\documentclass[a4paper,aps,prd,twocolumn,groupedaddress,amssymb,eqsecnum,showpacs,nofootinbib]{revtex4}
\usepackage{graphicx}
\usepackage{mathptmx}
\usepackage{bm}
\usepackage{times}

\begin{document}

%\preprint{CMU-08-nn}

\title{Unparticles and inflation}

\author{Hael Collins}
\email{hael@nbi.dk}
\affiliation{The Niels Bohr International Academy, The Niels Bohr Institute, 2100 Copenhagen \O, Denmark}
\author{R.~Holman}
\email{rh4a@andrew.cmu.edu}
\affiliation{Department of Physics, Carnegie Mellon University, 
Pittsburgh, Pennsylvania\ \ 15213}

\date{\today}

\begin{abstract}
We study some of the roles for unparticles in an inflationary universe.  Unparticles by themselves are not appropriate for generating the primordial perturbations since their power spectrum does not match what has been inferred from observations.  In fact, when the scaling dimension for the unparticles exceeds three-halves, the unparticle power spectrum diverges.  However, when a unparticle couples to an ordinary inflaton particle, loop corrections can produce a slight enhancement of the inflaton's power spectrum at longer wavelengths.  We examine these loop corrections from unparticles in some detail to learn how they scale in the wavelength of a perturbation and how they depend on the scaling dimension of the field.  
\end{abstract}


\maketitle

\section{Introduction}
\label{intro}

As our picture for nature reaches to ever smaller scales, it becomes more difficult to see new phenomena directly.  In the early days of bubble chamber experiments, a new particle could leave a distinctive track of its existence. In contrast, in current experiments the complexity of the products of a particle collision makes it far more challenging to understand what we are seeing without first viewing it through some sort of theoretical lens.  There is a danger that we might be blinded by our expectations, that it will be hard to see something entirely new unless we have already, to some measure, anticipated it.

This worry is illustrated by the example of {\it unparticles\/} \cite{georgi}.  Unparticles are scale invariant, or---not quite equivalently---conformal, fields that have a nonintegral scaling dimension.  As their name suggests, these fields lack many of the properties that normally characterize particle theories.  If they exist, they would produce rather different signatures in detectors from those of the more familiar particles that are usually considered.  Fortunately the scale invariance of an unparticle allows many of its interactions with ordinary particles to be treated simply and analytically.

Unparticles could have a role in cosmology too, particularly in the early universe \cite{astrounpart}.  This article looks at some of the ways they could affect the inflationary picture.  Most ambitiously, the inflaton---the field that drives the inflationary expansion---could itself be an unparticle.  We shall show that the conformal properties of unparticles, however, make them very unsuitable for this purpose.  Yet, even if they are not themselves responsible for inflation, unparticles can still produce interesting effects.  If they interact with an ordinary particle inflaton, the loop corrections from unparticles can enhance the power spectrum of the primordial perturbations produced by inflation.  This enhancement occurs whenever the scaling dimension of the unparticle field is less than three-halves and it is most pronounced at the large-wavelength end of the power spectrum.

This article begins in the next section by explaining how unparticles are incorporated into an isotropically expanding background.  Section~\ref{inflatons} shows why they are not especially good inflatons while Sec.~\ref{loops} examines in detail how they could influence the standard inflationary picture through radiative corrections.  The final section presents our conclusions and we leave a few simple loop calculations in flat space to an appendix, which also includes a description of how unparticles might be useful in regularizing conformal particle field theories.

\section{Preliminaries}
\label{prelim}

\subsection{Unparticles in a conformally flat universe}

An unparticle is nothing more than a scale-invariant field with a nontrivial scaling dimension, $d$.  The simplest and most familiar scalar field with these properties is not an unparticle at all, but is rather an ordinary massless particle, corresponding to the case $d=1$.  Once we go beyond integral values for $d$, however, the fields no longer have many of the properties that we associate with particles.  There are no isolated poles in the propagator and there is no longer a simple dispersion relation between the energy and the momentum carried by an unparticle.  Because of its scaling or conformal properties, an unparticle in a conformally flat space-time is closely related to one in flat space, so we shall first examine this relation in a little detail for the case of a conformal particle.

In a curved background, it is not enough to set the mass of a particle to zero to insure that the theory remains scale-invariant, since the background itself introduces another potential source for breaking it.  To restore the scale invariance, the field should be coupled conformally to the background.  Thus, for example, the $d=1$ prototypical scale-invariant field in a curved background is a massless, conformally coupled field,
\begin{equation}
S = \int d^4x\, \sqrt{-g}\, \Bigl[ 
{\textstyle{1\over 2}} g^{\mu\nu} \partial_\mu\varphi \partial_\nu\varphi 
- {\textstyle{1\over 12}} R \varphi^2 
\Bigr] . 
\label{mcc}
\end{equation}
As in flat space, a field of dimension $d=1$ is an ordinary particle theory, but other possibilities exist, which are genuine unparticles when $d>1$.\footnote{Unitarity \cite{ira}, in any event, requires $d\ge 1$.}  Using the conformal invariance of the theory provides a powerful constraint on how to extend our description of unparticles into an expanding background.  

Let us consider a massless, conformally coupled particle in an isotropically expanding background, expressed in conformally flat coordinates, 
\begin{equation}
ds^2 = a^2(\eta)\, \bigl[ d\eta^2 - d\vec x\cdot d\vec x \bigr] , 
\label{metric}
\end{equation}
in terms of which the scalar curvature becomes 
\begin{equation}
R = {6a^{\prime\prime}\over a^3} 
\equiv {6\over a^3} {\partial^2 a\over\partial\eta^2} . 
\label{Rvalue}
\end{equation}
If we expand the field in eigenmodes,
\begin{equation}
\varphi(\eta,\vec x) = \int {d^3\vec k\over (2\pi)^3}\, 
\Bigl[ \varphi_k(\eta) e^{i\vec k\cdot\vec x} a_{\vec k} 
+ \varphi_k^*(\eta) e^{-i\vec k\cdot\vec x} a_{\vec k}^\dagger \Bigr] , 
\label{phimodes}
\end{equation}
then the action implies that the modes obey the equation, 
\begin{equation}
\varphi_k^{\prime\prime} + 2 {a'\over a} \varphi_k 
+ {a^{\prime\prime}\over a}\varphi_k + k^2\varphi_k = 0 . 
\label{KGmode}
\end{equation}
Since the field is conformally coupled and the background here is itself conformally flat, we can immediately see that the modes are equivalent to the usual flat-space modes, up to a factor of the scale factor $a(\eta)$ raised to the rescaling dimension, which is in this case just $d=1$, 
\begin{equation}
\varphi_k(\eta) = {\chi_k(\eta)\over a(\eta)} ,
\label{confrescale}
\end{equation}
so that  
\begin{equation}
\chi_k^{\prime\prime} + k^2\chi_k = 0 . 
\label{KGchimode}
\end{equation}

Even though we may no longer have an explicit form for their action, the same principle applies to the unparticles.  If we denote an unparticle in flat space by $\chi(x)$ and the unparticle in a conformally flat space by $\sigma(x)$, then we can again write the latter as a simple rescaling of the former, 
\begin{equation}
\sigma(\eta,\vec x) = {\chi(\eta,\vec x)\over a^d(\eta)} , 
\label{relatingunparticles}
\end{equation}
determined by its rescaling dimension $d$.

\subsection{A short review of unparticles in flat space}

In a Lorentz-invariant theory, the propagator of any scalar field can be expressed in a Lehmann-K\" allen spectral form, 
\begin{eqnarray}
&&\!\!\!\!\!\!\!\!\!\!\!\!\!\!\!
\langle 0| T\bigl(\chi(x)\chi(y)\bigr) |0\rangle 
\label{LK} \\
&&= 
\int {d^4k\over (2\pi)^4}\, e^{-ik\cdot(x-y)} 
\int_0^\infty dM^2\, {i\rho(M^2)\over k_0^2 - |\vec k|^2 - M^2 + i\epsilon} . 
\nonumber
\end{eqnarray}
For an ordinary particle theory, $\rho(M^2)$ contains at least one pole at the physical mass of the field.  The field is then also renormalized so that the residue at this pole is $i$, which amounts to a renormalization condition on the theory.  In contrast, for an unparticle theory, we do not have any such conditions available to determine $\rho(M^2)$.  Instead, we appeal directly to the scale invariance of $\chi$ and the fact that its rescaling dimension is $d$ to fix $\rho(M^2)$ completely, up to a constant factor, $A_d$, 
\begin{equation}
\rho(M^2) = {A_d\over 2\pi} (M^2)^{d-2} .
\label{rhounpart}
\end{equation}
Unlike the particle case, we do not have a renormalization condition to set the normalization, $A_d$, of the unparticle; but a frequently used convention \cite{georgi} is to choose it to match with the phase space for $d$ massless particles, 
\begin{equation}
A_d = {16\pi^{5/2}\over (2\pi)^{2d}} 
{\Gamma(d+{1\over 2})\over \Gamma(2d)\Gamma(d-1)} 
= {d-1\over (16\pi^2)^{d-1}} {2\pi\over (\Gamma(d))^2} , 
\label{Addef}
\end{equation}
except that, of course, here $d$ is not an integer.

The integral over $M^2$ in the propagator converges as long as $1<d<2$; performing this integral then yields a simpler expression for the unparticle propagator, 
\begin{eqnarray}
&&\!\!\!\!\!\!\!\!\!\!\!\!\!\!\!
\langle 0| T\bigl(\sigma(x)\sigma(y)\bigr) |0\rangle 
\label{unpropflat2} \\
&&= 
\int {d^4k\over (2\pi)^4}\, e^{-ik\cdot(x-y)} 
{A_d\over 2}
{(-1)^{d-2}\over\sin\pi d} {i\over (k_0^2-|\vec k|^2+i\epsilon)^{2-d}} . 
\nonumber 
\end{eqnarray}
Notice that in the limit $d\to 1$, the $d-1$ factor in $A_d$ cancels the zero in $\sin\pi d$.  The propagator becomes that of a free, massless {\it particle\/}, correctly normalized; so in a sense, massless particles are unparticles too.  But we shall usually avoid this limit, since the some of the factors become singular while others vanish.

In an inflationary setting, where it might not be appropriate to treat the theory using an $S$-matrix formalism, it is usually more convenient to break the propagator into the two components of its time-ordering.  In flat space we have 
\begin{eqnarray}
&&\!\!\!\!\!\!\!\!\!\!\!\!\!
\langle 0| T\bigl(\chi(t,\vec x)\chi(t',\vec y)\bigr) |0\rangle 
\nonumber \\ 
&\!\!\!=\!\!\!& 
\int {d^3\vec k\over (2\pi)^3}\, e^{i\vec k\cdot(\vec x-\vec y)} 
\bigl[ 
\Theta(t-t')\, \tilde\Gamma_k^>(t,t') 
+\ \Theta(t'-t)\, \tilde\Gamma_k^<(t,t') 
\bigr] . 
\nonumber \\
&&
\label{propflat}
\end{eqnarray}
The functions 
\begin{eqnarray}
\tilde\Gamma_k^>(t,t') 
&\!\!\!=\!\!\!& 
\int d^3\vec x\, e^{-i\vec k\cdot(\vec x-\vec y)}\,
\langle 0| \chi(t,\vec x) \chi(t',\vec y) |0\rangle 
\nonumber \\
&\!\!\!=\!\!\!& \tilde\Gamma_k^<(t',t)
\label{wightdef}
\end{eqnarray}
are the {\it Wightman functions\/} for the unparticle field in flat space, written in their momentum representation.  Here we have used $k = |\vec k|$ to denote the magnitude of the spatial momentum, rather than the full four-momentum.  

Using the integrated form for the propagator in Eq.~(\ref{unpropflat2}), the unparticle Wightman function is 
\begin{equation}
\tilde\Gamma_k^>(t,t') = {A_d\over 2} {(-1)^{d-2}\over\sin\pi d} 
\int_{-\infty}^\infty {dk_0\over 2\pi}\, 
{i e^{-ik_0(t-t')}\over (k_0^2 - k^2 + i\epsilon)^{2-d}} ; 
\label{unWight}
\end{equation}
evaluating this integral \cite{gradshteyn} we find\footnote{To keep the expression a little more concise, we have not explicitly written the $i\epsilon$ in this expression; it is readily restored by replacing $k\to\sqrt{k^2-i\epsilon}$, as needed.} 
\begin{equation}
\tilde\Gamma_k^>(t,t') = {-i\over (8\pi^2)^{d-1}}
{\sqrt{\pi}\over 2\sqrt{2} } 
{1\over\Gamma(d)} 
\biggl( {t-t'\over k} \biggr)^{{3\over 2} - d} 
H_{d-{3\over 2}}^{(2)}[k(t-t')] . 
\label{unWightM}
\end{equation}
Here, $H_\nu^{(2)}(z)$ is one of the standard Hankel functions.  We have assumed that $t>t'$ in evaluating the momentum integral, which is automatically guaranteed by the accompanying $\Theta$-function. 

Although we have been reviewing the properties of unparticles in flat space, our results immediately generalize to an unparticle $\sigma$ in a conformally flat background, 
\begin{equation}
\langle 0 | T\bigl( \sigma(\eta,\vec x)\sigma(\eta',\vec y) \bigr) | 0\rangle 
= {\langle 0 | T\bigl( \chi(\eta,\vec x)\chi(\eta',\vec y) \bigr) | 0\rangle 
\over a^d(\eta) a^d(\eta')} , 
\label{confprop}
\end{equation}
which equally applies to the Wightman functions of the unparticle field too, $\Gamma_k(\eta,\eta')= a^{-d}(\eta)a^{-d}(\eta')\tilde\Gamma_k(\eta,\eta')$,
\begin{eqnarray}
\Gamma_k^>(\eta,\eta') 
&\!\!\!=\!\!\!& 
{-i\over (8\pi^2)^{d-1}}
{\sqrt{\pi}\over 2\sqrt{2} } 
{1\over\Gamma(d)} 
\biggl( {\eta-\eta'\over k} \biggr)^{{3\over 2} - d} 
\nonumber \\
&&
{1\over a^d(\eta) a^d(\eta')}
H_{d-{3\over 2}}^{(2)}[k(\eta-\eta')] , 
\label{unWightC}
\end{eqnarray}
and $\Gamma_k^<(\eta,\eta')=\Gamma_k^>(\eta',\eta)$.

Had we not integrated over the spectral parameter $M^2$ earlier, the Wightman function would have assumed a form that more explicitly shows its similarity to its flat space form,
\begin{eqnarray}
\Gamma_k^>(\eta,\eta') 
&\!\!\!=\!\!\!& {1\over a^d(\eta) a^d(\eta')} {A_d\over 2\pi}
\nonumber \\
&&
\int_0^\infty dM^2\, (M^2)^{d-2} {e^{-i(\eta-\eta')\sqrt{M^2+k^2-i\epsilon}}\over 2 \sqrt{M^2+k^2-i\epsilon}} . 
\qquad
\label{altWight} 
\end{eqnarray}
This form will be useful in the next section where we derive the power spectrum for an unparticle inflaton.

\section{Unparticles as inflatons}
\label{inflatons}

At a first glance, unparticles might appear to provide an interesting alternative for the inflaton itself, since the conformality of the field might be used to control some fine-tunings that occur for an ordinary inflaton.  In the usual inflationary picture, the effective mass of the inflaton must be quite small during the slowly rolling phase when the primordial perturbations are produced.  However, this case corresponds to a massless {\it minimally\/} coupled field, rather than a massless {\it conformally\/} coupled one, which is the natural starting point for an unparticle.  During inflation, the curvature changes rather slowly, so its coupling to the field effectively acts as a large mass term for a particle inflaton---the $d=1$ case---which produces a highly tilted power spectrum for the primordial perturbations.  Generalizing to the unparticle case, $d>1$, only worsens the shape of the power spectrum, as we shall learn in this section.

In inflation, the tiny primordial perturbations begin as the fluctuations of a quantum field.  The actual field responsible for producing the primordial perturbations is a combination of the inflaton and the scalar component of the metric fluctuations, but we shall simplify the picture somewhat by treating this combination as a pure unparticle.  The quantum fluctuations continually occur during the slowly rolling phase and are stretched to vast scales, whereupon they become essentially frozen into the space-time background.  The simplest measure of this pattern, which provides the initial input used to describe a variety of features of our universe, is the two-point function or equivalently its Fourier transform, the {\it power spectrum\/}, $P_k(\eta)$.  For an unparticle inflaton, these quantities are defined by 
\begin{equation}
\langle 0 | \sigma(\eta,\vec x) \sigma(\eta,\vec y) | 0\rangle 
= \int {d^3\vec k\over (2\pi)^3}\, e^{i\vec k\cdot(\vec x-\vec y)} 
{2\pi^2\over k^3} P_k(\eta) . 
\label{twopointdef}
\end{equation}
The power spectrum therefore has a simple relation to the unparticle Wightman function, as in Eq.~(\ref{altWight}), 
\begin{eqnarray}
P_k(\eta) 
&\!\!\!=\!\!\!& 
{k^3\over 2\pi^2}\Gamma_k^>(\eta,\eta) 
\nonumber \\
&\!\!\!=\!\!\!& 
{k^{2d}\over a^{2d}(\eta) } {A_d\over 8\pi^3} 
\int_0^\infty d\mu\, {\mu^{d-2} \over \sqrt{\mu+1}} ,
\label{powerwight} 
\end{eqnarray}
where we have rescaled the spectral parameter to produce a dimensionless integral, $\mu\equiv M^2/k^2$.  Notice that this integral diverges at the upper end of the integral unless $d<{3/2}$, so we find a narrower range of allowed values for the scaling dimension of the unparticle field.\footnote{Had we used the alternate form for the Wightman function given in Eq.~(\ref{unWightC}), it would have diverged as well as we took the limit, $\eta'\to\eta$, unless---as before---$d<{3\over 2}$.}  Otherwise, as long as $1<d<{3\over 2}$, we find
\begin{equation}
P_k(\eta) = {1\over (16\pi^2)^d} {4\over\sqrt{\pi} } 
{\Gamma({3\over 2}-d)\over\Gamma(d)}
\biggl( {k\over a(\eta) } \biggr)^{2d} . 
\label{powerunpart}
\end{equation}
For a de Sitter background, the scale factor is 
\begin{equation}
a(\eta) = - {1\over H\eta} , 
\label{adS}
\end{equation}
where $\eta$ runs from $-\infty$ to $0$, which gives a power spectrum, 
\begin{equation}
P_k(\eta) = {H^{2d}\over (16\pi^2)^d} {4\over\sqrt{\pi} } 
{\Gamma({3\over 2}-d)\over\Gamma(d)} (-k\eta)^{2d} . 
\label{powerunpartdS}
\end{equation}
The scaling in the wavenumber $k^{2d}$ does not at all match what has been observed; the actual primordial power spectrum appears to be nearly flat, which could only happen if $d\approx 0$.

\section{Unparticles in radiative corrections}
\label{loops}

Although we mentioned in the beginning that unparticles provide just one example of how nature might surprise us, they are actually not so exotic when compared with other possibilities.  Unparticles can even arise from a particle theory with the right properties.  As an example, consider a setting where at very high energies we have three sorts of particles:  the inflaton, some proto-unparticles, and a heavy particle of mass $M$ that interacts with the other two.  At energies below $M$, we can integrate out the heavy particles to generate interactions between the inflaton and the proto-unparticles, suppressed by some power of $1/M$.  If the proto-unparticles have a suitable infrared fixed point, at still lower energies the proto-unparticles will be replaced by an effective theory of unparticles with an associated dimensional transmutation scale $\Lambda$.  Assembling all of these ingredients together, the typical effective interaction between the inflaton, $\varphi$, and the unparticles, $\sigma$, will have the form, 
\begin{equation}
c \Lambda^{4 - r - s d} \biggl( {\Lambda\over M} \biggr)^n 
\varphi^r \sigma^s , 
\label{typical}
\end{equation}
where $c$ is a dimensionless coupling constant and $n$, $r$, and $s$ are some integers.  In an inflationary setting, the Hubble scale $H$ is a low energy scale, at least with respect to the dynamics that produced the unparticles.  This observation implies that we have the following hierarchy of scales, 
\begin{equation}
H < \Lambda < M . 
\label{hierarch}
\end{equation}

\subsection{Leading particle-unparticle interactions}

Even if the inflaton is not itself an unparticle, the presence of unparticles during inflation can still influence the pattern of primordial perturbations.  The leading influence will be through the most relevant, or lowest dimension, operators.  As before, let us denote the inflaton, now an ordinary particle, by $\varphi$ and assume that its potential remains invariant when we replace $\varphi$ with $-\varphi$.  The most important possible operator is the cubic interaction
\begin{equation}
c \Lambda^{2-d} \biggl( {\Lambda\over M} \biggr)^n\, \varphi^2\sigma . 
\label{cubic}
\end{equation}
Here, $c$ is a dimensionless coupling constant.  Although this operator is relevant, being $2+d$ dimensional, its effects can reasonably small, especially at the upper end of the range $1<d<2$ where it is almost marginal.  Then, even a small value of $n$ is usually sufficient to prevent the unparticles from overwhelming the usual flat power spectrum produced by the inflaton.  This section examines the leading indirect role that unparticles can have on the production of the primordial perturbations in inflation through their loop corrections.

The power spectrum for the inflaton is defined exactly as before,
\begin{eqnarray}
&&\!\!\!\!\!\!\!\!\!\!\!\!\!\!\!\!\!\!\!\!\!\!\!\!\!\!\!\!\!\!
\langle 0(\eta)|\varphi(\eta,\vec x)\varphi(\eta,\vec y)|0(\eta)\rangle 
\nonumber \\
&\!\!\!=\!\!\!& \int {d^3\vec k\over (2\pi)^3}\, e^{i\vec k\cdot(\vec x-\vec y)}\, 
\biggl[ {2\pi^2\over k^3} P_k(\eta) \biggr] , 
\label{loopdef}
\end{eqnarray}
except that now that we must go beyond the simplest prediction, the time-evolution of the states becomes important since it is affected by the unparticles.\footnote{Since the universe evolves only over a finite time interval, we use the Schwinger-Keldysh-Mahanthappa \cite{sk} formalism to solve the time-evolution of matrix elements.  A more detailed description of how this approach is applied to an inflationary setting is explained in \cite{effectivestate}.}

In the interaction picture, the evolution of a state is governed by Dyson's equation, 
\begin{equation}
|0(\eta) \rangle = T e^{-i\int_{\eta_0}^\eta d\eta'\, H_I(\eta')}\, |0\rangle , 
\label{dyson}
\end{equation}
where $|0\rangle$ denotes the initial state, $|0(\eta_0)\rangle$.  Similarly, its dual evolves as 
\begin{equation}
\langle 0(\eta)| = \langle 0|\, \biggl( T e^{-i\int_{\eta_0}^\eta d\eta'\, H_I(\eta')} \biggr)^\dagger ; 
\label{dysondag}
\end{equation}
notice that the conjugation effectively reverses time-ordering.  The interaction Hamiltonian is given by the operator at the beginning of the section, 
\begin{eqnarray}
H_I(\eta) 
&\!\!\!=\!\!\!& - c \Lambda^{2-d} \biggl( {\Lambda\over M} \biggr)^n 
\int d^3\vec x\, \sqrt{-g}\, \varphi^2\sigma 
\nonumber \\
&\!\!\!=\!\!\!& - c {\Lambda^{2-d}\over H^4\eta^4} 
\biggl( {\Lambda\over M} \biggr)^n \int d^3\vec x\, \varphi^2\sigma , 
\label{Hint}
\end{eqnarray}
which we have evaluated in de Sitter space.  

As before, we write the inflaton propagator in terms of its Wightman functions,
\begin{eqnarray}
&&\!\!\!\!\!\!\!\!\!\!\!\!\!
\langle 0| T\bigl(\varphi(\eta,\vec x)\varphi(\eta',\vec y)\bigr) |0\rangle 
\nonumber \\ 
&\!\!\!=\!\!\!& 
\int {d^3\vec k\over (2\pi)^3}\, e^{i\vec k\cdot(\vec x-\vec y)} 
\bigl[ 
\Theta(\eta-\eta')\, G_k^>(\eta,\eta') 
\nonumber \\
&&\qquad\qquad\qquad
+\ \Theta(\eta'-\eta)\, G_k^<(\eta,\eta') 
\bigr] , 
\label{dSpropflat}
\end{eqnarray}
which is the more convenient form when we would like to evaluate the time-evolution of an entire matrix element, rather than just a scattering matrix element.  Expanding the field in its operator eigenmodes,
\begin{equation}
\varphi(t,\vec x) = \int {d^3\vec k\over (2\pi)^3}\, 
\Bigl[ 
\varphi_k(t) e^{i\vec k\cdot\vec x} a_{\vec k} 
+ \varphi_k^*(t) e^{i\vec k\cdot\vec x} a_{\vec k}^\dagger 
\Bigr] , 
\label{phieigenmodes}
\end{equation}
produces a very simple form for the Wightman functions, 
\begin{equation}
G_k^>(\eta,\eta') = \varphi_k(\eta) \varphi_k^*(\eta') , 
\label{wightasU}
\end{equation}
with $G_k^<(\eta,\eta') = G_k^>(\eta',\eta)$.  The eigenmodes of a massive, minimally coupled scalar field obey the usual Klein-Gordon equation, 
\begin{equation}
\bigl[ \nabla^2 + m^2 \bigr] \varphi(\eta,\vec x) = 0, 
\label{KGgen}
\end{equation}
which for a purely de Sitter background implies
\begin{equation}
{d^2\varphi_k\over d\eta^2} - {2\over\eta} {d\varphi_k\over d\eta} 
+ \biggl( k^2 + {1\over\eta^2} {m^2\over H^2} \biggr) \varphi_k 
= 0 , 
\label{dSKG}
\end{equation} 
for the eigenmodes.  

The normalization of the modes is set by the equal-time commutation relation between $\varphi$ and its conjugate momentum, but to fix the solution completely requires making some assumption about the behavior of the modes.  The most typical assumption is to choose the maximally symmetric state that matches with the flat-space vacuum at infinitesimal intervals.\footnote{In flat space, small scales remain small, but that is not so in an expanding space-time, especially an inflating one.  So we ought---if we were considering the more general case---to be a little cautious in our assumptions about how nature behaves at extremely small scales \cite{transplanck}.}  This condition defines a state known as the Bunch-Davies vacuum \cite{bunch}, which in de Sitter space is given by
\begin{equation}
\varphi_k(\eta) = {\sqrt{\pi}\over 2} H \eta^{3/2} H_\nu^{(2)}(k\eta) , 
\label{bunchvac}
\end{equation}
where the index depends on the mass of the field,
\begin{equation}
\nu = \sqrt{ {9\over 4} - {m^2\over H^2} } . 
\label{nudef}
\end{equation}

For simplicity, and since the effective mass ought to be very small during the inflationary phase, we shall work in the limit of a perfectly massless inflaton, $m\to 0$ and $\nu\to {3\over 2}$, for which the modes, 
\begin{equation}
\varphi_k(\eta) = {H\over\sqrt{2}}
{(i-k\eta) e^{-ik\eta}\over k^{3/2}} , 
\label{bunchvacnomass}
\end{equation}
and the Wightman functions, 
\begin{equation}
G^>_k(\eta,\eta') = {H^2\over 2k^3} 
\bigl(1 + ik(\eta-\eta') + k^2\eta\eta'\bigr) e^{-ik(\eta-\eta')} , 
\label{bunchWight}
\end{equation}
simplify considerably.  Note, however, that a strictly massless field is not well behaved at long distances in de Sitter space.  It will be important to keep this fact in mind later as we examine the loop correction in detail.

With all of these ingredients now assembled, we can evaluate the effect of the unparticles on the power spectrum.  We shall assume that their influence is relatively small so that we can analyze the power spectrum perturbatively, 
\begin{equation}
P_k(\eta) = P_k^{\rm tree}(\eta) + P_k^{\rm loop}(\eta) + \cdots . 
\label{powerpert}
\end{equation}
As before, the tree-level term has a simple relation to the Wightman function, this time for a massless particle, 
\begin{equation}
P_k^{\rm tree}(\eta) = {k^3\over 2\pi^2} G_k^>(\eta,\eta) , 
\label{powerpert0}
\end{equation}
which in a de Sitter background becomes 
\begin{equation}
P_k^{\rm tree}(\eta) = {H^2\over 4\pi^2} \bigl(1 + k^2\eta^2 \bigr) . 
\label{powerpert0dS}
\end{equation}
The important modes for cosmology are those whose wavelength has been stretched well outside the Hubble horizon before the end of inflation, 
\begin{equation}
{k\over|a(\eta)|} \ll H(\eta) ; 
\label{latemodes}
\end{equation} 
in a de Sitter background, where $a^{-1}(\eta) = -H\eta$, this condition becomes 
\begin{equation}
|k\eta| \ll 1 ,
\label{dSlatemodes}
\end{equation} 
so the leading behavior of the power spectrum is a perfectly flat, scale-independent, spectrum,
\begin{equation}
P_k^{\rm tree}(\eta) \approx {H^2\over 4\pi^2} .
\label{zeropowerflat}
\end{equation}

\subsection{The loop}

\begin{figure}[!tbp]
\includegraphics{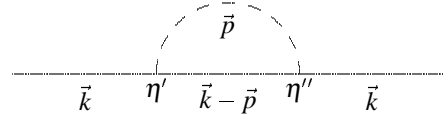}
\caption{A radiative correction from an unparticle to the two-point function of the inflaton.  In this picture, the inflaton is denoted by the solid line while the unparticle is denoted by the dashed line.\label{loop1}}
\end{figure}

For a cubic coupling of the form $\varphi^2\sigma$, the first influence of the unparticles is through the diagram shown in Fig.~\ref{loop1}, where we have chosen our loop momentum $\vec p$ so that $\vec k-\vec p$ runs through the internal inflaton line and $\vec p$ runs through the unparticle line.  The contribution from this diagram to the power spectrum is 
\begin{eqnarray}
P_k^{\rm loop}(\eta) 
&\!\!\!=\!\!\!& 
- {4c^2\over\pi^2} k^3 \Lambda^{2(2-d)} \biggl( {\Lambda\over M} \biggr)^{2n} 
\label{dPfrw} \\
&&\times
\int_{\eta_0}^\eta d\eta'\, a^4(\eta') 
\bigl\{ G_k^>(\eta,\eta') - G^<(\eta,\eta') \bigr\} 
\nonumber \\
&&\times
\int_{\eta_0}^{\eta'} d\eta^{\prime\prime}\, a^4(\eta^{\prime\prime}) 
\bigl\{ 
G_k^>(\eta,\eta^{\prime\prime}) 
L^>_k(\eta',\eta^{\prime\prime}) 
\nonumber \\
&&\qquad\qquad\qquad\quad 
-\ G_k^<(\eta,\eta^{\prime\prime}) 
L^<_k(\eta',\eta^{\prime\prime}) 
\bigr\} . \qquad
\nonumber 
\end{eqnarray}
Here the loop integrals are defined by
\begin{eqnarray}
L^>_k(\eta',\eta^{\prime\prime}) 
&\!\!\!=\!\!\!& 
\int {d^3\vec p\over (2\pi)^3}\, 
G_{|\vec p-\vec k|}^>(\eta',\eta^{\prime\prime}) 
\Gamma_p^>(\eta',\eta^{\prime\prime}) 
\nonumber \\
L^<_k(\eta',\eta^{\prime\prime}) 
&\!\!\!=\!\!\!& 
\int {d^3\vec p\over (2\pi)^3}\, 
G_{|\vec p-\vec k|}^<(\eta',\eta^{\prime\prime}) 
\Gamma_p^<(\eta',\eta^{\prime\prime}) , 
\qquad
\label{loopintdefs}
\end{eqnarray}
and the Wightman functions are given in Eq.~(\ref{unWightC}) and Eq.~(\ref{bunchWight}).    

The unparticle loop correction is rather complicated, so we first reexpress it as a few nested dimensionless integrals so that we can readily see how the correction $P_k^{\rm loop}(\eta)$ depends on the important physical scales.  For example, if we assume that $k\not= 0$, then we can define the conformal time in units of the wavenumber, 
\begin{equation}
x = - k\eta , 
\label{xdef}
\end{equation}
and similarly for each of the other internal or initial times that appear, 
\begin{equation}
x_0 = - k\eta_0, \quad x' = - k\eta' \quad\hbox{and}\quad 
x^{\prime\prime} = - k\eta^{\prime\prime} . 
\label{xdefs}
\end{equation}
We have included a sign in each of these dimensionless variables so that they are all positive, $x,x',x^{\prime\prime},x_0>0$. As we have assumed that $k$ does not vanish, which would correspond to an unobservable infinite mode, we can also rescale the loop momentum, 
\begin{equation}
q \equiv {p\over k} . 
\label{xidef}
\end{equation}
In terms of these dimensionless variables, the loop integral defined in Eq.~(\ref{loopintdefs}) becomes 
\begin{eqnarray}
L_k^>(x',x^{\prime\prime}) 
&\!\!\!=\!\!\!& - {i\over (8\pi^2)^{d+{1\over 2}}} {\pi^{3/2}\over\Gamma(d)} 
{H^{2d+2}\over k^3} 
\nonumber \\
&& 
x^{\prime\, d} x^{\prime\prime\, d} (x^{\prime\prime}-x')^{{3\over 2}-d}
\hat L^>(x',x^{\prime\prime}) 
\quad
\label{dimlessloop}
\end{eqnarray}
where
\begin{eqnarray}
\hat L^>(x',x^{\prime\prime}) 
&\!\!\!=\!\!\!& 
\int_0^\infty dq\, q^{d-{1\over 2}} 
H_{d-{3\over 2}}^{(2)}[q(x^{\prime\prime}-x')] 
\nonumber \\
&&
\biggl\{ 
\biggl[ {1\over |q-1|} - {ix'x^{\prime\prime}\over x^{\prime\prime}-x'} 
\biggr] e^{-i|q-1|(x^{\prime\prime}-x')} 
\nonumber \\
&&
-\ \biggl[ {1\over q+1} - {ix'x^{\prime\prime}\over x^{\prime\prime}-x'} 
\biggr] e^{-i(q+1)(x^{\prime\prime}-x')} 
\biggr\} , \qquad
\label{dimlessloopdef}
\end{eqnarray}
once we have integrated over all of the angular variables, using the integrals provided in Appendix~\ref{afewintegrals}.

Adding the external legs as well, we find the following expression for the unparticle loop correction, 
\begin{equation}
P_k^{\rm loop}(\eta) 
= {H^2\over 4\pi^2} {c^2\over\Gamma(d)} 
{ 2\sqrt{2\pi} \over (8\pi^2)^d}
\biggl( {\Lambda\over H} \biggr)^{4-2d} 
\biggl( {\Lambda\over M} \biggr)^{2n} 
\Pi(x) 
\label{P2scaling}
\end{equation}
where $\Pi(x)$ is the following, completely dimensionless, integral, 
\begin{eqnarray}
&&\!\!\!\!\!\!\!\!\!\!\!\!\! 
\Pi(x) 
\nonumber \\
&\!\!\!\equiv\!\!\!& 
\int_x^{x_0} {dx'\over x^{\prime\, 4-d}}\, 
\bigl[ (1+xx')\sin(x'-x) - (x'-x)\cos(x'-x) \bigr] 
\nonumber \\
&& 
\int_{x'}^{x_0} {dx^{\prime\prime}\over x^{\prime\prime\, 4-d}}\, 
\biggl\{
\bigl[ 
1 + i(x^{\prime\prime}-x)+xx^{\prime\prime} 
\bigr] 
e^{-i(x^{\prime\prime}-x)} 
{\hat L^>(x',x^{\prime\prime})\over (x^{\prime\prime}-x')^{d-{3\over 2}}}
\nonumber \\
&&\qquad\qquad
-\ \bigl[ 
1 - i(x^{\prime\prime}-x)+xx^{\prime\prime} 
\bigr] 
e^{i(x^{\prime\prime}-x)} 
{\hat L^<(x',x^{\prime\prime})\over (x'-x^{\prime\prime})^{d-{3\over 2}}} 
\biggr\} . 
\nonumber \\
&&
\label{Pidef} 
\end{eqnarray}
Before looking more closely at $\Pi(x)$, wherein lies all of the dependence on the conformal time and the wave number of the modes, we can already obtain a rough idea of the size of this correction by noting that the numerical factor, 
\begin{equation}
8.0\cdot 10^{-4} < {2\sqrt{2\pi}\over \Gamma(d) (8\pi^2)^d} < 6.3\cdot 10^{-2} \qquad\hbox{for}\quad d\in [1,2]
\label{pisandgammas}
\end{equation}
is actually quite small in the allowed range for the scaling dimension of the unparticle field.  From our hierarchy of scales, $H < \Lambda < M$, we should also have 
\begin{equation}
\biggl( {\Lambda\over M} \biggr)^n  
< \biggl( {H\over\Lambda} \biggr)^{2-d} 
\label{scalesuppress}
\end{equation}
which, given that $1<d<2$, requires only a very moderate value for $n$, especially when we include the numerical suppression already present in Eq.~(\ref{pisandgammas}).  The only remaining question is how the dimensionless part, $\Pi(x)$, depends on the initial and final times.

So far, aside from neglecting the mass of the inflaton, we have not made any other approximations.  However, there are two other useful limits that we can apply.  Earlier, when evaluating the tree-level power spectrum, we noted that the physically important modes for cosmology are those that have been stretched well outside the horizon by the end of inflation, $\eta$.  In a de Sitter background, for these modes $-k\eta$ is very small so we shall temporarily set $-k\eta = x\to 0$ in $\Pi(x)$.  A second limit, which is less general since we do not know much about how long any presumed inflationary epoch might have lasted or what might have preceded it, is to take $-k\eta_0 = x_0\to \infty$.  In these limits, the expression for the dimensionless integral simplifies enough that we can write the loop factors explicitly, 
\begin{widetext}
\begin{eqnarray}
\Pi(0) 
&\!\!\!=\!\!\!& 
\int_0^\infty dq\, q^{d-{1\over 2}} 
\int_0^\infty {dx'\over x^{\prime\, 4-d}}\, 
\bigl[ \sin(x') - x'\cos(x') \bigr] 
\int_{x'}^\infty {dx^{\prime\prime}\over x^{\prime\prime\, 4-d}}\, 
(x^{\prime\prime}-x')^{{3\over 2}-d}
\nonumber \\
&& 
\biggl\{
{1\over |q-1|} 
\biggl[
[ 1 + ix^{\prime\prime} ] 
e^{-i|q-1|(x^{\prime\prime}-x')-ix^{\prime\prime}} 
H_{d-{3\over 2}}^{(2)}[q(x^{\prime\prime}-x')] 
+ [ 1 - ix^{\prime\prime} ] 
e^{i|q-1|(x^{\prime\prime}-x')+ix^{\prime\prime}} 
H_{d-{3\over 2}}^{(1)}[q(x^{\prime\prime}-x')] 
\biggr]
\nonumber \\
&& 
- {1\over q+1} 
\biggl[
[1 + ix^{\prime\prime}]e^{-i(q+1)(x^{\prime\prime}-x')-ix^{\prime\prime}} 
H_{d-{3\over 2}}^{(2)}[q(x^{\prime\prime}-x')] 
+ [1 - ix^{\prime\prime}] 
e^{i(q+1)(x^{\prime\prime}-x')+ix^{\prime\prime}} 
H_{d-{3\over 2}}^{(1)}[q(x^{\prime\prime}-x')] 
\biggr]
\nonumber \\
&& 
- {ix'x^{\prime\prime}\over x^{\prime\prime}-x'} 
\biggl[
[1 + ix^{\prime\prime}] 
H_{d-{3\over 2}}^{(2)}[q(x^{\prime\prime}-x')] 
e^{-ix^{\prime\prime}} 
\Bigl[ 
e^{-i|q-1|(x^{\prime\prime}-x')} - e^{-i(q+1)(x^{\prime\prime}-x')} 
\Bigr]
\nonumber \\
&&\qquad\quad
-\ [1 - ix^{\prime\prime}] 
H_{d-{3\over 2}}^{(1)}[q(x^{\prime\prime}-x')] 
e^{ix^{\prime\prime}} 
\Bigl[ 
e^{i|q-1|(x^{\prime\prime}-x')} - e^{i(q+1)(x^{\prime\prime}-x')} 
\Bigr] 
\biggr]
\biggr\} . 
\label{Pi0def}
\end{eqnarray}
\end{widetext}

While this integral might appear still somewhat daunting, what is important experimentally is not its exact value---since we do not know anyway the strength of the coupling $c$ between the inflaton and the unparticle---but whether and how it diverges.  Any potential divergences can have one of two distinct origins.  First, there may simply be a standard short-distance divergence which corresponds to a need to renormalize the inflaton in the presence of unparticle interactions.  The second sort of divergence may result from the fact that we cannot completely neglect the mass of the inflaton or size of a mode at the end of inflation.  The possibility of a divergence as $x\to 0$ is the most interesting, since it means that the loop correction has a non-negligible dependence on $k$, which would provide a distinct observable signature.

Three of the regions within the range of integration in $\Pi(0)$ can be characterized as short-distance limits:  $q\to\infty$, short spatial separations, $x^{\prime\prime}\to x'$, coincident internal times, and $x'$, $x^{\prime\prime} \to\infty$, an arbitrarily early initial time.  This last case might not appear to be a short-distance limit, but we should remember that as we proceed into the past, physical lengths become ever smaller, and thus they are infinitesimally tiny in the infinite past.  Each of these limits is completely safe---that is, no divergences occur---as long as $d<2$.  We can verify this behavior by expanding the integrand in $\Pi(0)$ explicitly, but it is perhaps more instructive and simpler to look at the analogous calculation of the correction in flat space, which is done in Appendix~\ref{flatloops}.  This appendix shows that the loop correction in a Minkowski background is completely finite and does not require the renormalization of the inflaton that would be required for a particle loop.  At short distances, the curvature of de Sitter space should become negligible---we earlier chose our modes to be precisely those that matched with the flat-space modes in this same limit---so there should not be any short distance divergences in $\Pi(0)$ either.

The second group of divergences are long-distance ones:  $q\to 0$, $q\to 1$ and $x',x^{\prime\prime}\sim x\to 0$.  Because $q=p/k$ and because we have chosen the loop momentum so that $\vec p$ runs through the unparticle leg and $\vec k - \vec p$ runs through the inflaton leg, $q=0$ and $q=1$ correspond respectively to having a vanishing momentum in the virtual unparticle or the virtual inflaton.  As long as $d$ is positive, a soft unparticle does not produce any divergences.  However, a soft inflaton does yield a mild logarithmic divergence; near $q=1$ we have 
\begin{eqnarray}
\Pi(0) 
&\!\!\!=\!\!\!& 
\int_{1-\epsilon}^{1+\epsilon} dq\, {1\over |q-1|} 
\nonumber \\
&&
\int_0^\infty {dx'\over x^{\prime\, 4-d}}\, \bigl[ \sin(x) - x\cos(x) \bigr] 
\int_{x}^\infty {dx^{\prime\prime}\over x^{\prime\prime\, 4-d}}\, 
(x^{\prime\prime}-x')^{{3\over 2}-d}
\nonumber \\
&& 
\Bigl[
[ 1 + ix^{\prime\prime} ] e^{-ix^{\prime\prime}} 
H_{d-{3\over 2}}^{(2)}[(x^{\prime\prime}-x')] 
\nonumber \\
&&\quad
+ [ 1 - ix^{\prime\prime} ] e^{ix^{\prime\prime}} 
H_{d-{3\over 2}}^{(1)}[(x^{\prime\prime}-x')] 
\Bigr] 
\nonumber \\
&&
+ \cdots .
\label{qonediv} 
\end{eqnarray}
We mentioned earlier that a massless inflaton is not well defined in a de Sitter background, so this divergence should be cured by including a small mass $m$ for this field, so that here $\epsilon \sim m/k$ or 
\begin{equation}
\Pi(0) \sim \ln{m\over k} + \cdots .
\label{qonelog} 
\end{equation}

There remains only the case where $x'$ and $x^{\prime\prime}$ are small.  The lower limit---actually the upper limit in the conformal time---of the $x'$ integral is not exactly zero but is rather $x=-k\eta$.  In this limit the leading behavior of the integrand is 
\begin{eqnarray}
\Pi(x) 
&\!\!\!=\!\!\!& {1\over 3} {4\sqrt{2}\over 2^d\Gamma(d-{1\over 2})} 
\int_0 dq\, q^{2d-2} 
\biggl\{ 
{1\over |q-1|} - {1\over q+1} 
\biggr\} 
\nonumber \\
&&
\int_x dx'\, x^{\prime\, d-1} 
\int_{x'} dx^{\prime\prime}\, x^{\prime\prime\, d-4} 
+ \cdots 
\label{smallxdiv} 
\end{eqnarray}
and we discover that time-integrals scale as 
\begin{equation}
\int_x dx'\, x^{\prime\, d-1} 
\int_{x'} dx^{\prime\prime}\, x^{\prime\prime\, d-4} 
\propto x^{2d-3} = (-k\eta)^{2d-3} , 
\label{smallxscale} 
\end{equation}
which would diverge when $d<{3\over 2}$ if we took $k\eta\to 0$.

So by examining the asymptotic behavior of the unparticle loop correction, $\Pi(x)$, we have discovered that it can have some important scaling when the dimension of the field lies within the range $1<d<{3\over 2}$.  If we let $\Pi_i$ denote a dimensionless integral which is completely finite as $m\to 0$ and $k\eta \to 0$, then we can write $\Pi(x)$ in a form where all of its leading, observable dependence on the physical parameters of the theory is made explicit, 
\begin{equation}
\Pi(x) = \Pi_0 + \Pi_1\, \ln{m\over k} 
+ \Pi_2\, (-k\eta)^{2d-3} + \Pi_3\, (-k\eta)^{2d-3}\ln{m\over k} 
\label{PiasPis}
\end{equation}
so that the correction to the power spectrum due to an unparticle becomes, 
\begin{eqnarray}
P_k^{\rm loop}(\eta) 
&\!\!\!=\!\!\!& 
{H^2\over 4\pi^2} {c^2\over\Gamma(d)} 
{ 2\sqrt{2\pi} \over (8\pi^2)^d}
\biggl( {\Lambda\over H} \biggr)^{4-2d} 
\biggl( {\Lambda\over M} \biggr)^{2n} 
\nonumber \\
&&
\Bigl\{ 
\Pi_0 + \Pi_1\, \ln{m\over k} + \Pi_2\, (-k\eta)^{2d-3} 
\nonumber \\
&&
+ \Pi_3\, (-k\eta)^{2d-3}\ln{m\over k} 
\Bigr\} . 
\label{P2scalingfull}
\end{eqnarray}

\section{Conclusion}
\label{conclude}

An inflaton that is also an unparticle does not appear to produce a realistic power spectrum.  The power spectrum for the primordial perturbations inferred from the precise measurements of the cosmic microwave background radiation \cite{wmap} is nearly flat, with perhaps a little more power at long wavelengths than at smaller ones.  Unitarity \cite{ira} already requires that $d>1$, so that for scaling dimensions where $d<{3\over 2}$, where the power spectrum scales as $k^{2d}$, unparticles do not produce a flat spectrum.  Beyond $d>{3\over 2}$, the power spectrum becomes poorly defined since the Wightman function diverges at coincident times for such scaling dimensions.  Because of its conformal properties, the flat-space propagator for an unparticle is closely related to that of a conformally flat background; so this divergence applies equally well to unparticles in flat space.

An unparticle is much more useful as an additional ingredient in the standard inflationary picture.  As an example, the loop correction produced when an unparticle interacts with a scalar inflaton particle tends to add to the long wavelength end of the power spectrum.  We analyzed the loop correction due to a cubic coupling between the inflaton and an unparticle; we found that its leading behavior scales as $(k\eta)^{2d-3}$.  For modes stretched well outside the horizon during inflation, for which $k\eta\to 0$, this term becomes negligible for $d>{3\over 2}$ but below three-halves, it enhances the power in the large-wavelength modes.

Although we have been primarily interested in the properties of unparticles in an inflationary setting, along the way we have learned a few more general things about them that also apply to flat space.  Aside from the divergence in the Wightman function just mentioned, the behavior of unparticles is in some ways tamer than that of ordinary particles.  Simple loop corrections that contain only particles usually diverge.  These divergences arise form the large momentum region of the loop integral and they are canceled by renormalizing some aspect of the theory.  Unparticles in loops tend to be free of these short-distance divergences, as is shown for a few examples in the following appendix.  This observation suggests that unparticles can be useful as a method for regularizing conformal particle theories, particularly for massless fermions---provided their unparticle analogues exist---since such a conformal regularization does not require breaking or extending any of the space-time symmetries.

\begin{acknowledgments}

\noindent
This work was supported in part by DOE grant No.~DE-FG03-91-ER40682, by the EU FP6 Marie Curie Research and Training Network ``UniverseNet'' (MRTN-CT-2006-035863), and by the Niels Bohr International Academy.

\end{acknowledgments}

\appendix

\section{Loops in flat space}
\label{flatloops}

The short-distance behavior of unparticles---at least as virtual particles---is much milder than that of an ordinary scalar particle.  To illustrate this observation, we look at two simple, similar loop amplitudes for unparticles, evaluated in flat space and using the standard $S$-matrix formalism.  As before, we denote the flat-space unparticle field by $\chi$ and the ordinary---now massive ($m$)---particle field by $\varphi$.

\subsection{Two particles and one unparticle}

\begin{figure}[!tbp]
\includegraphics{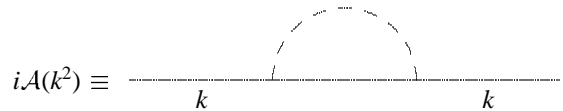}
\caption{A loop correction to the particle $\varphi$ propagator (solid line) from an unparticle $\sigma$ (dashed line) from a $\varphi^2\sigma$ interaction.\label{loop2}}
\end{figure}

Let us first choose the same coupling that we have been considering in an inflationary setting, 
\begin{equation}
c\Lambda^{2-d} \biggl( {\Lambda\over M} \biggr)^n \varphi^2\chi ,
\label{cubicflat}
\end{equation}
and calculate the loop amplitude associated with its correction to the $\varphi$-field propagator.  The diagram for this amplitude is shown in Fig.~\ref{loop2} and yields 
\begin{eqnarray}
{\cal A}(k^2) 
&\!\!\!=\!\!\!& c^2 \Lambda^{2(2-d)} \biggl( {\Lambda\over M} \biggr)^{2n}
{1\over (16\pi^2)^d} {\Gamma(1-d)\over\Gamma(d)} 
\nonumber \\
&&\times 
\int_0^1 dx\, \biggl[ {xm^2 - x(1-x)k^2 - i\epsilon\over 1-x} \biggr]^{d-1} , 
\end{eqnarray}
once we have integrated over the loop momentum.  As a technical point, the calculation proceeds most simply if we begin with the unparticle propagator in the form given in Eq.~(\ref{LK}), introduce the Feynman parameter integral ($dx$) {\it before\/} integrating over the spectral parameter $M^2$, and lastly integrate over the loop momentum.  The nonintegral scaling dimension of the field naturally avoids poles in the $\Gamma$-functions that result from this last integral.  As a consequence, unparticles when coupled as above do not require any renormalization of the particle field.  In the particle limit, $d\to 1$, the $\Gamma(1-d)$ factor produces the standard divergence.  

Note that the integrand contains a mild singularity at $x=1$; however, in the range $1<d<2$, this singularity is completely integrable, being of the form
\begin{equation}
\sim \int^1 dx\, {1\over (1-x)^{d-1}} . 
\label{intsing}
\end{equation}
Thus, for example, when the particle is on shell, $k^2=m^2$, 
\begin{eqnarray}
{\cal A}(m^2) 
&\!\!\!=\!\!\!& {c^2\Lambda^2\over (16\pi^2)^d} 
\biggl( {m\over\Lambda} \biggr)^{2(d-1)}
\biggl( {\Lambda\over M} \biggr)^{2n}
\nonumber \\
&&\times 
{\Gamma(1-d) \Gamma(2-d) \Gamma(2d-1)\over\Gamma(d) \Gamma(d+1)} . 
\end{eqnarray}

If we would like this correction to be ``natural,'' in the sense that it is not significantly larger than the mass scale of the particle, ${\cal A}\le m^2$, then $n$ needs to be sufficiently large to satisfy 
\begin{equation}
\biggl( {\Lambda\over M} \biggr)^{n}
< \biggl( {m\over\Lambda} \biggr)^{2-d} . 
\label{nreq}
\end{equation}

\subsection{Unparticle regularization}

Unparticles can be of some use even when we are only interested in a particle theory, as long as we have some massless particles around.  As we just saw, the nonintegral scaling dimension avoids the usual short-distance divergences in the loop.  We can therefore use an unparticle as a way to regulate a massless particle, while preserving the underlying $3+1$ dimensional Lorentz invariance of the theory.\footnote{This technique resembles the method of ``Analytic Renormalization'' proposed in \cite{speer}.}  

For example, starting with a massless particle $\chi$ coupled to a massive one, $\varphi$, by $c\Lambda\varphi^2\chi$, evaluating the loop in $4-2\delta$ dimensions yields 
\begin{equation}
{\cal A}_{\rm dr} = {c^2\Lambda^2\over 16\pi^2} \biggl[ 
{1\over\delta} - \gamma + \ln 4\pi 
- \int_0^1 dx\, \ln {xm^2 - x(1-x)k^2 - i\epsilon\over\mu^2} 
\biggr] . 
\label{Adr}
\end{equation}
Comparing with an unparticle coupling $c\Lambda \mu^{1-d}\varphi^2\chi$, the amplitude in the small $\delta$ limit---where now $d=1+\delta$---is 
\begin{eqnarray}
{\cal A}_{\rm un} 
&\!\!\!=\!\!\!& {c^2\Lambda^2\over 16\pi^2} \biggl[ 
- {1\over\delta} - 2\gamma - 1 + 2\ln 4\pi 
\nonumber \\
&&\qquad
- \int_0^1 dx\, \ln {xm^2 - x(1-x)k^2 - i\epsilon\over\mu^2} 
\biggr] ; 
\label{Aup}
\end{eqnarray}
and we see that the finite, physical parts of the amplitudes are exactly the same.

\subsection{One particle and two unparticles}

\begin{figure}[!tbp]
\includegraphics{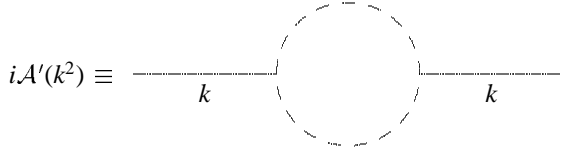}
\caption{A loop correction to the particle $\varphi$ propagator (solid line) from an unparticle $\sigma$ (dashed line) from a $\varphi\sigma^2$ interaction.\label{loop3}}
\end{figure}

We conclude with one last example, where this time the cubic coupling is 
\begin{equation}
c'\Lambda^{3-2d} \biggl( {\Lambda\over M} \biggr)^{n'} \varphi\sigma^2 . 
\label{cubicflat2}
\end{equation}
The calculation proceeds as before so we shall be a bit briefer.  The amplitude associated with the unparticle loop shown in Fig.~\ref{loop3} is 
\begin{equation}
{\cal A'}(k^2) = {c^{\prime\, 2} \Lambda^{2(3-2d)}\over (16\pi^2)^{2d-1}}
\biggl( {\Lambda\over M} \biggr)^{2n'}
 {\Gamma(2-2d)\over\Gamma(2d)} 
(-k^2-i\epsilon)^{2d-2} . 
\end{equation}
Notice that we still have the divergence at $d=1$ as well as a new divergence when $d={3\over 2}$, where $\Gamma(2-2d)\to\infty$.

When the external particle is on-shell, $k^2=m^2$, ``naturalness'' in this instance, ${\cal A}' \le m^2$, requires 
\begin{equation}
\biggl( {\Lambda\over M} \biggr)^{n'}
< \biggl( {m\over\Lambda} \biggr)^{3-2d} . 
\label{nPreq}
\end{equation}

\section{Angular integrals}
\label{afewintegrals}

In the loop integral, defined in Eq.~(\ref{loopintdefs}), occur several angular integrals of the general form, 
\begin{equation}
\int_{-1}^1 dz\, {e^{-i|\vec p-\vec k|(\eta'-\eta^{\prime\prime})}\over 
|\vec p-\vec k|^n} 
\label{cosintN}
\end{equation}
where $z\equiv \cos\theta$ is related to the angle between the loop momentum $\vec p$ and the external momentum $\vec k$.  We have also assumed that $\eta' > \eta^{\prime\prime}$, which is enforced by the time-ordering implicit in the limits of the internal conformal time integrals for the overall expression for the first correction to the two-point function.  The first ($n=1$) of these integrals is straightforward enough, 
\begin{eqnarray}
\int_{-1}^1 dz\, {e^{-i|\vec p-\vec k|(\eta'-\eta^{\prime\prime})}\over 
|\vec p-\vec k|} 
&\!\!\!=\!\!\!& 
{e^{-i|p-k|(\eta'-\eta^{\prime\prime})}\over ipk(\eta'-\eta^{\prime\prime})} 
\nonumber \\
&&
-\ {e^{-i(p+k)(\eta'-\eta^{\prime\prime})}\over ipk(\eta'-\eta^{\prime\prime})} , 
\label{cosint1}
\end{eqnarray}
while the other two can be thereafter generated from it, 
\begin{eqnarray}
\int_{-1}^1 dz\, {e^{-i|\vec p-\vec k|(\eta'-\eta^{\prime\prime})}\over 
|\vec p-\vec k|^2} 
&\!\!\!=\!\!\!& 
{{\rm Ei}\bigl(1, i|p-k|(\eta'-\eta^{\prime\prime}) \bigr)\over pk} 
\nonumber \\
&&
-\ {{\rm Ei}\bigl(1, i(p+k)(\eta'-\eta^{\prime\prime}) \bigr)\over pk} 
\label{cosint2}
\end{eqnarray}
and 
\begin{eqnarray}
&&\!\!\!\!\!\!\!\!\!\!\!\!\!\!\!\!\!\!\!\!\!\!\!\!\!\!\!\!\!\!\!\!\!\!
\int_{-1}^1 dz\, {e^{-i|\vec p-\vec k|(\eta'-\eta^{\prime\prime})}\over 
|\vec p-\vec k|^3} 
\nonumber \\
&\!\!\!=\!\!\!& - {i(\eta'-\eta^{\prime\prime})\over pk} \bigl[ 
{\rm Ei}\bigl(1, i|p-k|(\eta'-\eta^{\prime\prime}) \bigr)
\nonumber \\
&&\qquad\qquad\quad
-\ {\rm Ei}\bigl(1, i(p+k)(\eta'-\eta^{\prime\prime}) \bigr) \bigr]
\nonumber \\
&&
+\ {e^{-i|p-k|(\eta'-\eta^{\prime\prime})}\over pk|p-k|}
- {e^{-i(p+k)(\eta'-\eta^{\prime\prime})}\over pk(p+k)} . 
\label{cosint3}
\end{eqnarray}

Our convention here for the definition of the exponential integral is that used by the {\sc maple} program, which is 
\begin{equation}
{\rm Ei}(1,x) = \int_1^\infty d\zeta\, {e^{-x\zeta}\over\zeta} . 
\label{expint}
\end{equation}

\end{document}